\documentclass[graybox]{svmult}

\usepackage{makeidx}         % allows index generation
\usepackage{graphicx}        % standard LaTeX graphics tool
\usepackage{multicol}        % used for the two-column index
\usepackage[bottom]{footmisc}% places footnotes at page bottom

\usepackage{cite}
\usepackage{epsfig}
\usepackage{graphicx}
\usepackage{amsmath}
\usepackage{amssymb}
\usepackage{ulem}
\usepackage{mdwlist}
\usepackage{color}

\usepackage{newtxtext}       % 
\usepackage{newtxmath}       % selects Times Roman as basic font
\usepackage{cite}

\makeindex             % used for the subject index
\def\b0{\beta_0}
\newcommand{\ep}{\varepsilon}

\begin{document}

\title*{{\footnotesize \sf DESY 19--096, DO--TH 19/09, SAGEX-2019-13}\\ 
Three loop heavy quark form factors and their asymptotic behavior}
\titlerunning{Heavy quark form factors}
\author{J. Ablinger$^1$, J. Bl\"umlein$^2$, P. Marquard$^2$, N. Rana$^{2,3}$ and C. Schneider$^1$}
\institute{\email{Narayan.Rana@mi.infn.it} 
\at $^1$RISC, Johannes Kepler University, Altenbergerstra{\ss}e 69, A-4040, Linz, Austria 
\at $^2$DESY, Platanenallee 6, D-15738 Zeuthen, Germany
\at $^3$INFN, Sezione di Milano, Via Celoria 16, I-20133 Milano, Italy}
% \and Name of Second Author \at Name, Address of Institute \email{name@email.address}}
\maketitle

% \abstract*{}

\vspace{-3cm}

\abstract{
A summary of the calculation of the color--planar and complete light quark contributions to the massive 
three--loop form factors is presented. Here a novel calculation method for the Feynman integrals
is used, solving general uni--variate first order factorizable systems of differential equations. We also 
present predictions for the asymptotic structure of these form factors. 
}

\vspace{-0.4cm}
%-----------------------------------------------------------------------------------------------------------------
\section{Introduction}
%-----------------------------------------------------------------------------------------------------------------
\vspace{-0.1cm}
\noindent
The detailed description of top quark pair production to high perturbative order is of importance in
various respects, including precision studies of QCD, the measurement of the top-quark mass and other
of its properties, and in the search for effects from potential physics beyond the Standard Model.
The heavy quark form factors act as the basic building block of the related observables. In a series of 
publications \cite{Bernreuther:2004ih,Bernreuther:2004th,Bernreuther:2005rw,Bernreuther:2005gw}, 
two--loop QCD contributions of these form factors for vector, axial-vector, scalar and pseudo-scalar currents, 
were first computed. In an independent calculation in \cite{Gluza:2009yy}, the ${\cal O}(\varepsilon)$ terms 
were included for the vector form factors, where $\varepsilon$ is the dimensional regularization parameter in 
$D=4-2 \varepsilon$ space-time dimensions. Later in \cite{Ablinger:2017hst}, two-loop QCD contributions up to ${\cal 
O}(\varepsilon^2)$ for all these form factors were obtained.

At the three-loop level, the color-planar contributions to the vector form factors were obtained in \cite{Henn:2016tyf,Henn:2016kjz} 
and the complete light quark contributions in \cite{Lee:2018nxa}. We have computed both the color-planar and complete light 
quark contributions to the three-loop form factors for the axial-vector, scalar and pseudo-scalar currents in 
\cite{Ablinger:2018yae} and for the vector current in \cite{Ablinger:2018zwz}, which are the subject of the first part of this 
article. In Ref.~\cite{Ablinger:2018zwz}, we have presented a detailed description of the method which we have used to 
obtain the master integrals in this case. The method is generic to compute any first order factorizable and uni-variate 
system of differential equations. In a parallel calculation, the same results have been obtained in \cite{Lee:2018rgs}. 

Amplitudes for hard scattering processes in QCD provide a clear insight into underlying principles such as 
factorization or the universality of infrared (IR) singularities. In the case of massless QCD amplitudes, a plethora of work 
\cite{Catani:1998bh,Sterman:2002qn, Becher:2009cu, Gardi:2009qi} has been performed to understand the structure of IR 
divergences which is due to the interplay of the soft- and collinear dynamics. In the case of two parton amplitudes, i.e. 
the form factors, the IR structure is more prominent. The interplay of the soft and collinear anomalous dimensions building up 
the singular structure of the massless form factors has first been noticed in \cite{Ravindran:2004mb} at two-loop order and has 
been later established at three-loop order in \cite{Moch:2005tm}. The generalization of this universal structure to the case 
of massive form factors is also of interest. First steps were taken in \cite{Mitov:2009sv} in the asymptotic limit i.e. in 
the limit where the quark mass is small compared to the center of mass energy, followed by the proposition of a factorization theorem 
\cite{Penin:2005eh, Mitov:2006xs, Becher:2007cu} in the asymptotic limit. Finally, in \cite{Becher:2009kw}, a general 
solution was presented following a soft-collinear effective theory approach. While, the solution in \cite{Becher:2009kw} 
provides the structure of IR poles for the exact computation, the study of the Sudakov behavior in the asymptotic limit 
also elucidates the logarithmic behavior for the finite contributions. Following the method proposed for massless form 
factors in \cite{Ravindran:2005vv, Ravindran:2006cg}, we have performed a rigorous study in \cite{Blumlein:2018tmz} 
in the asymptotic limit to obtain all the poles and also all logarithmic contributions to finite pieces of the three-loop 
heavy quark form factors for vector, axial-vector, scalar and pseudo-scalar currents. 
A similar study has been performed in \cite{Ahmed:2017gyt} obtaining the poles for the vector form factor.
In the second part of this article, 
we summarize the contents of Ref.~\cite{Blumlein:2018tmz}.

\vspace{-0.4cm}
%-----------------------------------------------------------------------------------------------------------------------------
\section{Heavy quark form factors}
%-----------------------------------------------------------------------------------------------------------------------------
\vspace{-0.1cm}
\noindent
We consider a virtual massive boson of momentum $q$, which can be a vector ($V$), an axial-vector ($A$), a scalar ($S$) or a 
pseudo-scalar ($P$), decaying into a pair of heavy quarks of mass $m$, color $c$ and $d$ and momenta $q_1$ and $q_2$, at a 
vertex $X_{cd}$, where $X_{cd} = \Gamma^{\mu}_{V,cd}, \Gamma^{\mu}_{A,cd}, \Gamma_{S,cd}$ and $\Gamma_{P,cd}$.
The general forms of the amplitudes are
%-----------------------------------------------------------------------------------------------------------------------------
\begin{align}
\bar{u}_c (q_1) \Gamma_{V,cd}^{\mu} v_d (q_2) &\equiv
  -i \bar{u}_c (q_1) \Big[ \delta_{cd}
 v_Q \Big( \gamma^{\mu} ~F_{V,1} 
         + \frac{i}{2 m} \sigma^{\mu \nu} q_{\nu}  ~ F_{V,2}  \Big) \Big] v_d (q_2) , 
\nonumber\\
%%%%%
\bar{u}_c (q_1) \Gamma_{A,cd}^{\mu} v_d (q_2) &\equiv
  -i \bar{u}_c (q_1) \Big[ \delta_{cd}
 a_Q \Big( \gamma^{\mu} \gamma_5 ~F_{A,1} 
         + \frac{1}{2 m} q_{\mu} \gamma_5  ~ F_{A,2}  \Big) \Big] v_d (q_2) , 
\nonumber\\
%%%%%
\bar{u}_c (q_1) \Gamma_{S,cd} v_d (q_2) &\equiv
  -i \bar{u}_c (q_1) \Big[ \delta_{cd}
 s_Q \Big( \frac{m}{v} (-i) ~F_{S}  \Big) \Big] v_d (q_2) , 
\nonumber\\
%%%%%
\bar{u}_c (q_1) \Gamma_{P,cd} v_d (q_2) &\equiv
  -i \bar{u}_c (q_1) \Big[ \delta_{cd}
 p_Q \Big( \frac{m}{v} (\gamma_5) ~F_{P}  \Big) \Big] v_d (q_2) \,.
\end{align}
%-----------------------------------------------------------------------------------------------------------------------------
Here $\bar{u}_c (q_1)$ and $v_d (q_2)$ are the bi--spinors of the quark and the anti--quark, respectively, with 
$\sigma^{\mu\nu} = \frac{i}{2} [\gamma^{\mu},\gamma^{\nu}]$. $v_Q, a_Q, s_Q$ and $p_Q$ are the Standard Model (SM) coupling 
constants for the 
vector, axial-vector, scalar and pseudo-scalar, respectively. $v = (\sqrt{2} G_F)^{-1/2}$ denotes the SM vacuum expectation 
value of the Higgs field, with the Fermi constant $G_F$. For more details, see \cite{Ablinger:2017hst}.
The form factors can be obtained from the amplitudes by multiplying appropriate projectors \cite{Ablinger:2017hst} and performing 
the trace over the color and spinor indices.

\vspace{-0.6cm}
%-----------------------------------------------------------------------------------------------------------------------------
\subsection{Details of the computation}
%-----------------------------------------------------------------------------------------------------------------------------
\vspace{-0.2cm}
\noindent
The computational procedure is described in detail in Ref.~\cite{Ablinger:2017hst}. 
The Feynman diagrams are generated using {\tt QGRAF} \cite{Nogueira:1991ex}. The packages {\tt Q2e/Exp} 
\cite{Harlander:1997zb,Seidensticker:1999bb}, {\tt FORM} \cite{Vermaseren:2000nd, Tentyukov:2007mu}, and {\tt Color} 
\cite{vanRitbergen:1998pn} are used in the computation. By decomposing the dot products among the loop and external 
momenta into the combination of inverse propagators, each Feynman diagram can be expressed in terms of a linear combination of a 
large 
set of scalar integrals. These integrals are related to each other through integration-by-parts identities (IBPs)
\cite{Chetyrkin:1981qh,Laporta:2001dd}, and are reduced to 109 master integrals (MIs) by using the package {\tt Crusher} 
\cite{CRUSHER}. 

We apply the method of differential equations \cite{Kotikov:1990kg,Remiddi:1997ny,Henn:2013pwa,Ablinger:2015tua}
to calculate the master integrals. The method and the corresponding algorithm is presented in detail 
in \cite{Ablinger:2018zwz}.\footnote{For a review on the computational methods of loop integrals in quantum field theory, 
see \cite{Blumlein:2018cms}.}
The principal idea of this method is to obtain a set of differential equations of the MIs by performing differentiation
with respect to the variable $x$, with $q^2/m^2 = - (1-x)^2/x$  and then to use the IBP relations on the output to obtain  a linear 
combination of MIs for each 
differentiated integral for general bases. One obtains a $n \times n$ system of coupled linear differential equations for $n$ 
master integrals ${\cal I}$
%-----------------------------------------------------------------------------------------------------------------------------
\begin{equation}\label{Equ:InputSystem}
 \frac{d}{dx} {\cal I} = {\cal M ~ I + R}.
\end{equation}
%-----------------------------------------------------------------------------------------------------------------------------
Here the $n \times n$ matrix ${\cal M}$ consists out of entries from the rational function field ${\mathbb K}(D,x)$ 
(or equivalently from ${\mathbb K}(\ep,x)$) where ${\mathbb K}$ is a field of characteristic $0$. The inhomogeneous part ${\cal R}$ 
contains MIs which are already known. In simple cases, ${\cal R}$ turns out to be just the null vector. 
The first step to solve such a coupled system of differential equations
is to find out whether the system factorizes to first order or not. Using the 
package {\tt Oresys} \cite{ORESYS}, based on Z\"urcher's algorithm \cite{Zuercher:94} and applying a corresponding solver
\cite{SOLVER,Ablinger:2018zwz}
we have first confirmed that the present system is indeed first order factorizable in $x$-space. 

Without the need to choose a special basis, we solve the system in terms of iterated integrals over 
whatsoever alphabet, cf.~Ref.~\cite{Ablinger:2018zwz} for details. To proceed, we first arrange the differential equations
in such a manner that it appears in upper block-triangular form. Then, we compute the integrals block-by-block
starting from the last in the arrangement.
While solving for each block, say of order $m \times m$, 
the differential equations are solved order by order
in $\varepsilon$ successively, starting at the leading pole terms, $\propto 1/\varepsilon^3$ for our case. 
The successive solutions in $\varepsilon$ also contribute to the inhomogeneities in the next order. 
We use the 
package {\tt Oresys} \cite{ORESYS}, based on Z\"urcher's algorithm \cite{Zuercher:94} to uncouple the differential equations. 
At each order in $\ep$, $l$ inhomogeneous ordinary 
differential equations are obtained, where $1\leq l \leq m$. The orders of these differential equations are
$m_1, \ldots, m_l$ such that $m_1 + \cdots + m_l = m$.
We have solved these differential equations using the method of variation of constant.
In our case, the spanning alphabet is
%-----------------------------------------------------------------------------------------------------------------------------
\begin{equation}
\label{eq:LETT}
\frac{1}{x},~~
\frac{1}{1-x},~~
\frac{1}{1+x},~~
\frac{1}{1-x+x^2},~~
\frac{x}{1-x+x^2},
\end{equation}
%-----------------------------------------------------------------------------------------------------------------------------
i.e.~the usual harmonic polylogarithms (HPLs) \cite{Remiddi:1999ew} and their cyclotomic extension (CHPL) \cite{Ablinger:2011te}.
While integration over a letter is a straightforward algebraic manipulation, often $k$-th power of a letter, $k \in \mathbb{N}$, 
appears which needs to be transformed to the letters of (\ref{eq:LETT}) by partial integration. The other $m-l$ solutions are 
immediately obtained from the former solutions. 
The constants of integration are determined using boundary conditions in the low energy limit i.e. at $x = 1$. 
The boundary values for the HPLs and CHPLs give rise to the respective constants in the limit $x \rightarrow 1$, i.e. the multiple 
zeta values (MZVs) \cite{Blumlein:2009cf} and the cyclotomic constants \cite{Ablinger:2011te}.
The computation is performed by intense use of {\tt HarmonicSums}
\cite{Vermaseren:1998uu,Blumlein:1998if,Ablinger:2011te,Ablinger:2014rba, Ablinger:2010kw, Ablinger:2013hcp, Ablinger:2013cf, Ablinger:2014bra}, 
which uses the package {\tt Sigma} 
\cite{Schneider:sigma1,Schneider:sigma2}. Finally, all the MIs
have been checked numerically using {\tt FIESTA} \cite{Smirnov:2008py, Smirnov:2009pb, Smirnov:2015mct}.

%------------------------------------------------------------------------------------------------------------------

\vspace{-0.6cm}
\subsection{Ultraviolet renormalization and universal infrared structure}
\vspace{-0.2cm}
\noindent
We perform the ultraviolet (UV) renormalization of the form factors in a mixed scheme. The heavy quark mass and wave function 
have been renormalized in the on-shell (OS) renormalization scheme. The strong coupling constant has been renormalized using 
the $\overline{\rm MS}$ scheme, by setting the universal factor $S_\varepsilon = \exp(-\varepsilon (\gamma_E - \ln(4\pi))$ 
for each loop order to one at the end of the calculation.

The required renormalization constants are already well known and are denoted by 
$Z_{m, {\rm OS}}$ \cite{Broadhurst:1991fy, Melnikov:2000zc,Marquard:2007uj,Marquard:2015qpa,Marquard:2016dcn}, 
$Z_{2,{\rm OS}}$ \cite{Broadhurst:1991fy, Melnikov:2000zc,Marquard:2007uj,Marquard:2018rwx} and 
$Z_{a_s}$ \cite{Tarasov:1980au,Larin:1993tp}, with $a_s = \alpha_s/(4\pi)$,
for the heavy quark mass, wave function and strong coupling constant, respectively. 
$Z_{2,{\rm OS}}$ and $Z_{a_s}$ are multiplicative, while the renormalization of massive fermion 
lines has been taken care of by properly considering the counter terms.
For the scalar and pseudo-scalar currents, the presence of the heavy quark mass in the Yukawa coupling
employs another overall mass renormalization constant, which also has been performed in the OS renormalization scheme.

The universal behavior of IR singularities of the massive form factors was first
investigated in \cite{Mitov:2006xs} considering the high energy limit. Later in \cite{Becher:2009kw},
a general argument was provided to factorize the IR singularities as a multiplicative renormalization
constant as
%-------------------------------------------------------------------------------------------------------------
\begin{equation}
 F_{I} = Z (\mu) ~ F_{I}^{\mathrm{fin}} (\mu)\, ,
\end{equation}
%-------------------------------------------------------------------------------------------------------------
where $F_{I}^{\text{fin}}$ is finite as $\ep \rightarrow 0$. The renormalization group equation (RGE) for $Z(\mu)$ 
is constrained by the massive cusp anomalous dimension \cite{Grozin:2014hna,Grozin:2015kna}. 

\vspace{-0.6cm}
%-------------------------------------------------------------------------------------------------------------
\subsection{Checks of the results}
%-------------------------------------------------------------------------------------------------------------
\vspace{-0.2cm}
% We propose an algorithm to solve any first order factorizable and univariate system of differential equations.
% We apply this to obtain the master integrals appearing in the
% color--planar and complete light quark contributions to the massive form factors and finally we compute the form factors
% which are necessary ingredients to precision study of many observables related to top quark.
%
To perform checks, we have maintained the gauge parameter $\xi$ to first order and have thus obtained a partial check on gauge 
invariance. Fulfillment of the chiral Ward identity gives another strong check on our calculation.

Considering $\alpha_s$--decoupling appropriately, we obtain the universal IR structure for all the UV renormalized 
results, confirming again the universality of IR poles. Also, in the low energy limit, the magnetic vector form factor 
produces the anomalous magnetic moment of a heavy quark which we cross check with \cite{Grozin:2007fh} in this limit.
Finally, we have compared our results with those of Ref.~\cite{Henn:2016tyf, Lee:2018nxa, Lee:2018rgs}, which have been 
computed using partly different methods. Both results agree.

\vspace{-0.4cm}
%-------------------------------------------------------------------------------------------------------------
\section{Asymptotic behavior of massive form factors}
%-------------------------------------------------------------------------------------------------------------
\vspace{-0.1cm}
\noindent
We consider from now on only the renormalized electric form factor ($F_V$) for the vector current and the renormalized scalar 
form factor ($F_S$), in the asymptotic limit. All other massive form factors either agree to one of them or vanish in this limit.
To start with, we write down a Sudakov type integro-differential equation \cite{Collins:1980ih} for a function 
$\hat{F}_I \left(a_s(\mu), \frac{Q^2}{\mu^2}, \frac{m^2}{\mu^2}, \ep \right)$ in the asymptotic limit as follows
%--------------------------------------------------------------------------------------------------------
\begin{equation} \label{eq:kge}
 \mu^2 \frac{\partial}{\partial \mu^2} \ln \hat{F}_{I}\left(\frac{Q^2}{\mu^2},\frac{m^2}{\mu^2}, a_s,\ep\right) 
= \frac{1}{2} \left[ K_{I}\left(\frac{m^2}{\mu^2},a_s,\ep\right) + 
G_{I}\left(\frac{Q^2}{\mu^2},a_s,\ep\right) \right]\,,
\end{equation}
where $I=V,S$ only. 
%--------------------------------------------------------------------------------------------------------
Here $\hat{F}_I$ contains all logarithmic behavior and singular contributions of the respective form factor.
As evident from the functional dependence, $K_I$ incorporates the contributions from the quark mass $m$ and does not 
depend on the kinematic invariants, while $G_I$ contains the information of the process. 
Along with the evolution of the strong coupling constant, Eq.~\eqref{eq:kge}, and the renormalization group (RG) invariance
of $\hat{F}$, individual solutions for $K_I$ and $G_I$ are provided as follows
%--------------------------------------------------------------------------------------------------------
\begin{align} \label{eq:solnKG}
 K_{I} &=
 K_{I} \Big( a_s (m^2), 1, \ep \Big) - \int_{\frac{m^2}{\mu^2}}^{1}  \frac{d \lambda}{\lambda} A_q \Big( a_s (\lambda \mu^2) \Big), 
\nonumber\\
 G_{I} &=
 G_{I} \Big( a_s (Q^2), 1, \ep \Big) + \int_{\frac{Q^2}{\mu^2}}^{1}  \frac{d \lambda}{\lambda} A_q \Big( a_s 
(\lambda 
\mu^2) \Big).
\end{align}
%--------------------------------------------------------------------------------------------------------
Here $A_q$ denotes the quark cusp anomalous dimension. 
$K_{I} ( a_s (m^2), 1, \ep )$ and $G_{I} ( a_s (Q^2), 1, \ep)$ are initial conditions arising while solving 
the RG equations.
Using Eq.~\eqref{eq:solnKG}, one can solve Eq.~\eqref{eq:kge} to obtain $\hat{F}_I$, from which the form factors 
can be obtained through the following matching relation
%--------------------------------------------------------------------------------------------------------
\begin{equation}  \label{eq:ffexp}
{F}_{I} \Big( a_s, \frac{Q^2}{\mu^2}, \frac{m^2}{\mu^2}, \ep \Big) 
= {\cal C}_I(a_s,\ep) \hat{F}_{I} \Big( a_s, \frac{Q^2}{\mu^2}, \frac{m^2}{\mu^2}, \ep \Big)\,.
\end{equation}
%--------------------------------------------------------------------------------------------------------
The solutions for $\hat{F}_I$ up to four-loop are presented in Ref.~\cite{Blumlein:2018tmz}. 
At each order in $a_s$, say $n$, the solution consists of $A_I^{(n)}$, $K_I^{(n)}$ and $G_I^{(n)}$, the expansion coefficients
of $A_I$, $K_{I} ( a_s (m^2), 1, \ep )$ and $G_{I} ( a_s (Q^2), 1, \ep)$, respectively, and lower order terms.

In the massless quark form factor, the soft ($f_q$) and collinear ($B_q$) anomalous dimensions govern the infrared structure in 
the form $\gamma_q = B_q + \frac{f_q}{2}$. Intuitively, in the massive case, $\gamma_q$, along with similar contributions 
($\gamma_Q$) from the heavy quark anomalous dimension, will control the singular structure. Hence, it is suggestive to write
%--------------------------------------------------------------------------------------------------------
\begin{equation}
 K_I^{(n)} = - 2 ( \gamma_q^{(n)} + \gamma_Q^{(n)} - \gamma_I^{(n-1)} ) \,.
\end{equation}
%--------------------------------------------------------------------------------------------------------
The anomalous dimension $\gamma_I^{(n-1)}$ \cite{Broadhurst:1991fy,Vermaseren:1997fq,Gracey:2000am,
Melnikov:2000zc,Marquard:2007uj,Luthe:2016xec,Baikov:2017ujl} arises due to renormalization of the current.
Note that the power of each term $\gamma^{n}$ indicates the series expansion in $a_s$. For $\gamma_I$, the contribution 
is of the same order also, however we denote it by $(n-1)$ to match with general notation of \cite{Vermaseren:1997fq}.
The other finite functions $G_I^{(n)}$ contain the information on the process through its dependence on $Q^2$. Hence, 
it is similar to the one in case of massless form factors \cite{Ahmed:2014cla,Ahmed:2014cha}
%--------------------------------------------------------------------------------------------------------
\begin{equation} \label{eq:gin}
 G_I^{(n)} = 2 (B_q^{(n)} - \gamma_I^{(n-1)}) + f_q^{(n)} + C_I^{(n)} + \sum_{k=1}^{\infty} \ep^k g_{I}^{n,k}. 
\end{equation}
%--------------------------------------------------------------------------------------------------------
Given the structural similarities, $C_I^{(n)}$ and $g_{I}^{n,k}$ are the same as in the massless cases. All the required 
anomalous dimensions, except $\gamma_Q$, are known from different computations. On the other hand, $\gamma_Q$ can be obtained 
from the non-logarithmic contribution of the massive cusp anomalous dimension in the asymptotic limit. With all the ingredients, 
we obtain the full singular contributions and all logarithmic contributions to the finite part for vector and scalar form factors 
in the asymptotic limit. The non-logarithmic part of the finite piece gets contributions from the matching function ${\cal C}_I$ 
which can only be obtained by an exact computation. Using our results of \cite{Ablinger:2018zwz}, we obtain the color--planar 
and complete light quark contributions for ${\cal C}_I^{3,0}$.

\vspace{-0.4cm}
%--------------------------------------------------------------------------------------------------------
\section{Conclusion}
%--------------------------------------------------------------------------------------------------------
\vspace{-0.1cm}
In the first part, we have summarized the computational details to obtain the color--planar and 
complete light quark contributions to the three--loop heavy quark form factors along with a new method to solve uni-variate 
first order factorizable systems of differential equations. The system is solved in terms of iterative integrals 
over a finite alphabet of letters. Finally, we have computed all the corresponding contributions to the massive
three-loop form factors for vector, axial--vector, scalar and pseudo--scalar currents, which play an important role in 
the phenomenological study of the top quark. We then have studied the asymptotic behavior of these form factors. 
A Sudakov type integro--differential equation can be written down for the massive form factors and along with 
the study of RGE, we have obtained all the logarithmic contributions of the finite part of the vector and scalar 
form factors.

\noindent
{\bf Acknowledgment.}
This work was supported in part by the Austrian Science Fund (FWF) grant
SFB F50 (F5009-N15), by the bilateral project DNTS-Austria 01/3/2017 (WTZ
BG03/2017), funded by the Bulgarian National Science Fund and OeAD (Austria), by the EU TMR network SAGEX
Marie Sklodowska-Curie grant agreement No. 764850 and COST action CA16201:
Unraveling new physics at the LHC through the precision frontier.

%
% \section*{Appendix}
% \addcontentsline{toc}{section}{Appendix}
%
%

\vspace{-0.5cm}

% \bibliography{main}
% \bibliographystyle{MYutphysM}

\providecommand{\href}[2]{#2}\begingroup\raggedright\endgroup

\end{document}